# Atom-light-correlated quantum interferometer with memory-induced phase comb

Wenfeng Huang[1]†, Xinyun Liang[1]†, Jie Zhao[1], Zeliang Wu[1], Keye Zhang[1], Chun-Hua Yuan[1], Yuan Wu[1], Bixuan Fan[3]*, Weiping Zhang[2,4,5,6]* , Liqing Chen[1,3]*

[1]State Key Laboratory of Precision Spectroscopy, Department of Physics and Electronic Science, East China Normal University, Shanghai 200062, China.

[2]School of Physics and Astronomy, and Tsung-Dao Lee Institute, Shanghai Jiao Tong University, Shanghai 200240, China.

[3]College of Physics and Communication Electronics, Jiangxi Normal University, Nanchang 330022, China

[4]Shanghai Branch, Hefei National Laboratory, Shanghai 201315, China.

[5]Collaborative Innovation Center of Extreme Optics, Shanxi University, Taiyuan, Shanxi 030006, China.

[6]Shanghai Research Center for Quantum Science, Shanghai 201315, China.

* Corresponding authors: fanbixuan@jxnu.edu.cn; wpz@sjtu.edu.cn; lqchen@phy.ecnu.edu.cn;

† : These authors contributed equally to this work.

**Abstract:** Precise phase measurements by interferometers are crucial in science for detecting subtle changes, such as gravitational waves. However, phase sensitivity is typically limited by the standard quantum limit (SQL)[1-3] with uncorrelated particles $N$. This limit can be surpassed using quantum correlations[4-8], but achieving high-quality correlations in large systems is challenging[9-11]. Here, we propose and demonstrate an atom-light hybrid quantum interferometry whose sensitivity is enhanced beyond the SQL with atom-light quantum correlation and newly developed phase comb superposition via atomic-memory-assisted multiple quantum amplification. Finally, a phase sensitivity beyond the SQL of up to 8.3±0.2 dB is achieved, especially at $N=4\times10^{13}$/s, resulting in both atomic and optical phase sensitivities of $6\times10^{-8}$ rad/$\sqrt{\text{Hz}}$. This technique can advance sensitive quantum measurements in various fields.

Interferometers are the most precise instruments for extracting phase sensing information and are widely applied in gyroscopes for deep-space inertial guidance[12,13] and tests involving general relativity[14,15], ground and orbiting detection of gravitational waves [16-18], and imaging[19-23]. The phase sensitivity $\left(\delta\varphi = \langle\Delta\hat{O}\rangle/\left|\partial\langle\hat{O}\rangle/\partial\varphi\right|\right)$ is the core evaluation parameter of an interferometer and is determined by the noise $\langle\Delta\hat{O}\rangle$ and the slope $\left|\partial\langle\hat{O}\rangle/\partial\varphi\right|$ of the interference output, where $\hat{O}$ is the measurable operator of the interferometer output port. An interferometer operated with $N$ uncorrelated particles has a fundamental sensitivity limit scaled by $1/\sqrt{N}$, known as the standard quantum limit (SQL), which can improve slowly with increasing $N$ in principle. However, in most advanced interferometers for gravitational wave detection, gyroscopes, gravity, and bioimaging, phase sensitivity cannot be improved by continuously increasing $N$ due to the noise intrusion caused by a large $N$. Thus, quantum interferometry that can overcome the SQL in large $N$ systems is urgently needed for highly sensitive measurements.

To date, phase sensitivity beyond the SQL has been achieved by establishing quantum correlations between the two interference arms to contribute an enhancement factor $K$ to phase sensitivity as $\delta\varphi \propto 1/K\sqrt{N}$. This is accomplished either by enhancing the slope or by reducing the noise, as shown in several pioneering experiments ranging from photons [24-26] to Bose-Einstein condensates[27-29]. The best achievements are beyond the SQL by 15.6 dB in an $N$=26400 atom interferometer[27] and by 4.1 dB in an $N\sim10^{14}$ optical interferometer [25]. Quantum interferometry in large-N systems can provide further sensitivity breakthroughs in the currently available most sensitive measurement systems. However, these sensitive breakthroughs are extremely difficult owing to limitations in generating the quantum correlations of the large enhancement factors in large-$N$ systems[30-33], their fragility once produced[34-37], and the challenge of combining quantum correlation with other quantum technology[38-43].

Here, we propose and demonstrate an atom-light hybrid quantum interferometry technique by utilizing an atom-light quantum correlation and integrating the first developed phase comb superposition. Nonlinear Raman amplification is implemented in an $^{87}$Rb atomic vapor as a quantum splitter to generate a quantum correlation between the atomic arm and the optical arm. Atom-light quantum correlation can enhance the phase sensitivity surpassing the SQL. Phase-comb superposition occurs in two interference arms via multiple quantum amplification by combining an atomic memory function with quantum correlation, further enhancing the sensitivity beyond SQL.

**Phase-comb-enhanced interferometry**

As shown in Fig.1a, atom-light hybrid quantum interferometry was achieved using forward and backward Raman amplification processes as wave splitting and recombination of the quantum-correlated atomic arms and optical arms. A strong continuous wave $W$ spontaneously excites the vapor to an atomic collective spin state $S_a$ and generates a signal field $S_L$ through forward Raman amplification. $S_a$ and $S_L$, which serve as the atomic and optical interference arms, respectively, are quantum correlated. Specifically, their intensity difference is squeezed, and their phases are conjugated. A phase shift $\varphi$ is introduced into $S_L$, resulting in $S_L(\varphi)$ via mirror M. Then, $W$ and $S_L(\varphi)$ are reflected back to allow $\varphi$ to be memorized in $S_a$, resulting in $S_a(\varphi)$ via backward Raman amplification, with Raman gain G. This Raman amplification is a quantum amplification due to the quantum correlation between $S_a$ and $S_L$. This process is the first loop (leftmost panel) in Fig. 1b. Then, $S_a(\varphi)$ remains in the atomic vapour and serves as the initial state of subsequent forward Raman amplification via forward $W$ in the second loop. Here, for the first time, the atomic arm performs quantum amplification and phase memory functions. The entire process is described by the following Hamiltonian (see the Supplementary Materials. III):

$$\widehat{H} \simeq -\eta e^{i[\sin(\varphi)\kappa_s+\zeta]t}\hat{S}_a^\dagger\hat{S}_L^\dagger + H.C.$$  (1)

where the Raman strength $\eta \propto \frac{A_W}{\Delta}$, with $A_W$ and $\Delta$ being the amplitude and detuning of the $W$ field, respectively, and the Raman strength results in the Raman gain $G$ defined as follows: $G = \cosh(\eta t)$. $\sin(\varphi)\kappa_s$ is the phase shift term stored by the atomic spin, where $\kappa_s$ is the decay rate of the $S_L$ during the feedback loop. The H.C. term represents the Hermitian conjugate, and t is the time. The ac-Stark shift $\zeta \propto \frac{|A_W|^2}{\Delta}$ introduces a phase offset $\delta$ to $S_a$. Due to this structure, the atomic arm is cyclically amplified and memorizes a series of phase shifts $\varphi$ over time in a round-robin manner until the system reaches a steady state when the generation rate of atomic spin is equal to the atomic decay rate. This process involves a multiple quantum amplification plus phase memory. Each amplification is based on the output of the previous amplification, carries one more phase shift, and is memorized in the atomic arm. The final atomic spin arm achieves a superposition state upon accumulation of the phase shifts, constituting a phase comb with comb teeth $\varphi_n$. The corresponding optical arm $S_L$ also carries a set of phase shifts via the forward Raman wave splitting process. The atomic arm and optical arm in the steady state are defined as follows (see the Supplementary Materials. IV):

$$S_a = \sum_{n=1}^{J} G_n e^{i\varphi_n} S_L^*(0) \qquad (2)$$

$$S_L = \sum_{n=1}^{J} G_n' e^{i\varphi_n'} S_L(0), \qquad (3)$$

where the phase comb tooth $\varphi_n = n\varphi + \delta_n$, with $\delta_n$ being the phase offset in the $n$-th Raman amplification with gain factor $G_n$. The phase comb tooth $\varphi_n'$ in optical arm ($\varphi_n' = n\varphi + \delta_n'$) corresponds to the phase of $S_L$ after feedback, and $G_n'$ is the Raman gain for $S_L$. Both $\varphi_n'$ and $\varphi_n$ are related to phase shift $\varphi$. However, the phase offsets ($\delta_n$, $\delta_n'$) caused by ac-Stark effect are different. The intensities of $S_a$ and $S_L$ are greatly magnified from vacuum fields to high-gain regimes via multiple amplifications, thus generating a large number of atomic spin $S_a$ and $S_L$ photons and guaranteeing high absolute phase sensitivity. $S_L$ and $S_a$ are two-mode quantum correlated, and each arm has phase comb superposition. Both arms contribute to the final interference output $S_f$ with the following intensity:

$$I_{S_f} \propto \sum_{n=1}^{J} F_n \cos(n\varphi + \Lambda_n), \tag{4}$$

where $\Lambda_n$ is the phase offset from collective interference, whose value gradually increases as the loop number $n$ increases. The variation in $\Lambda_n$ causes successive shifting of the phase peak of the interference. Thus, $S_f$ is gradually amplified and deviates from the symmetric cosine shape used in the traditional interferometer to a sawtooth shape as $n$ increases, as shown in Fig. 1b. The sawtooth fringe has a much larger slope than the cosine fringe. The phase offset $\Lambda_n$ plays a significant role in achieving the sawtooth interference fringe. A larger $\Lambda_n$ correlates to a greater slope. The $F_n$ term is the intensity of each phase comb tooth, including the backward Raman amplification gain contributing to slope enhancement. Therefore, dual slope enhancements exist as follows: backward amplification and the sawtooth shape enhancement due to phase comb superposition. Thus, the slope of the current interferometer is much larger than that of a laser interferometer with the same phase-sensitive particle number $N$, whose best phase sensitivity is the SQL. The noise for a coherent amplifier increases simultaneously as the slope is amplified, preventing the sensitivity from exceeding the SQL. However, quantum correlation and phase comb superposition ensure dual slope enhancements while the noise remains unchanged. Thus, the phase sensitivity can be enhanced to surpass the SQL.

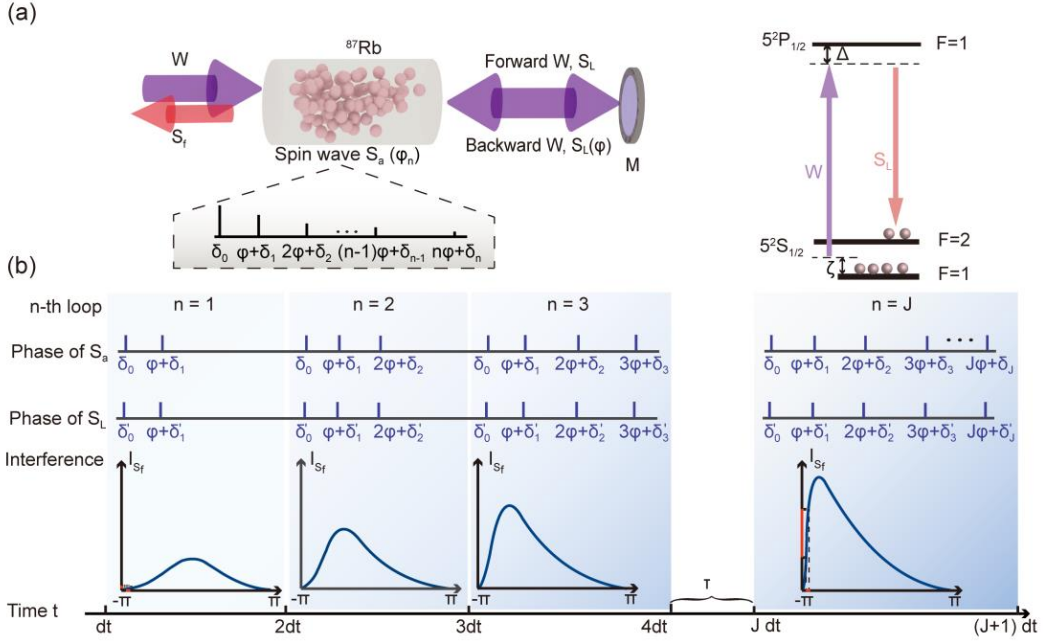

**Fig. 1 Phase-comb-enhanced atom–light hybrid quantum interferometry. (a)** Scheme and the energy level transitions. A forward continuous wave $W$ enters $^{87}$Rb atomic vapor, producing an atomic spin $S_a$ and a forward signal field $S_L$. $S_L$ is emitted from the cell and undergoes a phase shift, resulting in $S_L(\varphi)$, upon reflection by a mirror (M). The $W$ and $S_L(\varphi)$ beams are returned to the atomic cell so that the phase shift $\varphi$ can be recorded in $S_a$ via backward Raman amplification. Then, atomic spins stay in atomic vapor as the intital state for subsequent forward Raman amplification. This process is operated periodically within the atomic memory. The final interference arms $S_a$ and $S_L$ both carry a set of phase shifts, forming a phase comb. $S_f$ is the optical interference output. $\Delta$: the detuning frequency of the W field. $\zeta$: ac-Stark frequency shift. $\delta_n$ ($n=0\sim J$): $\zeta$-induced phase shift in the $n$-th loop with $J$ as the loop number. **(b)** Phase accumulation and interference fringe evolution over time. dt: time duration, i.e., the travel time of the $W$ and $S_L$ fields out of and back into the atomic vapor.

**Sensitivity to optical phase and atomic phase**

To demonstrate the operation of the interferometer, the fringe intensity and atomic phase shift are measured. The setup and energy levels are shown in Fig. 2a. The intensity of the $S_f$ field in the steady state (Fig. 2b, red solid line) changes with the optical phase shift $\varphi$ in the form of a sawtooth-shaped fringe due to phase comb superposition. This result clearly shows the optical-phase-sensitive property of current interferometry. The slope of $S_f$, $\left|\frac{\partial \langle \bar{N} \rangle}{\partial \varphi}\right|$, is much larger, especially at $\varphi = 0 \sim 0.5\pi$, than that of the laser interferometer, whose fringe has a symmetric cosine shape (Fig. 2b, green dashed line). This is one advantage of current interferometry. When the atomic spin couples an atomic phase shift, denoted as $\theta_A$, as $W$ and $S_L$ exit and enter the atomic system, the atomic arm $S_a$ in the steady state is $S_a = \sum_{n=1}^{J} G_n e^{i(\varphi_n + \theta_A)} S_L^*(0)$. Therefore, the intensity of the interference output $S_f$, $I_{S_f} \propto \sum_{n=1}^{J} F_n \cos(n\varphi + \Lambda_n + \theta_A)$ is also sensitive to the atomic phase $\theta_A$. Fig. 2c shows $S_f$ over the phase shift with and without external atomic phase modulation $\theta_A$ to illustrate the above point. The atomic phase is obtained by using probe light P to illuminate the atoms, with $\theta_A \propto \frac{I_p \tau}{\Delta'}$, where $I_p$ and $\Delta'$ represent the power and detuning of the probe light, respectively. $\tau \sim 1\mu s$ is the delay time of the 300m long fibre between the two Raman processes in the forward and backward directions. The phase shift of $S_f$ linearly changes with respect to the probe power and is inversely proportional to the detuning frequency $\Delta'$ (Fig. 2d), demonstrating an atomic-phase-sensitive property as well. The current sawtooth-shaped interferometer is thus an atom-light hybrid interferometer.

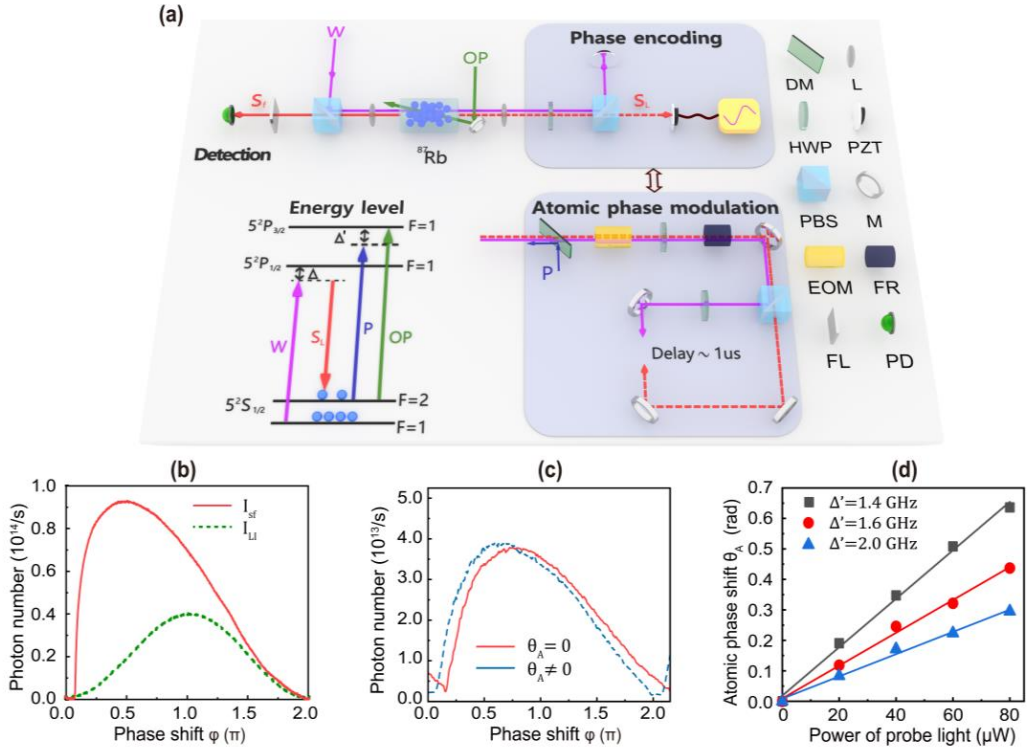

**Fig. 2 Experimental operation of the atom-light hybrid interferometer with a sawtooth interference fringe.** (**a**) Experimental setup. OP: optical pumping field; P: probe field; PBS: polarizing beam splitter; HWP: half-wave plate; PZT: piezoelectric transducer; L: len; M: mirror; DM: dichroic mirror; EOM: electro-optical modulator; FL: filter; FR: Faraday rotator. $\Delta'$: detuning of the probe field. (**b**) Interference fringe $S_f$ (red solid line) for optical phase shift $\varphi$ scanned from 0 to $2\pi$. Green dashed line represents the interference output of a laser interferometer(LI) with the same number of phase-sensing particles as the red solid line, $N=4\times10^{13}$/s. (**c**) Interference fringe $S_f$ with (blue dashed line) and without (red solid line) an atomic phase shift $\theta_A$ induced by the ac-Stark shift effect of probe field $P$. The detuning frequency of the probe field is $\Delta'=1.6$ GHz. The power of the probe field is 50 μW. (**d**) Measured atomic phase shift $\theta_A$ as a function of the power of the probe field at different detuning frequencies $\Delta'$. In figures (b-d), the power of the $W$ field is 250 μW with a red detuning of $\Delta=1.1$ GHz and the power of the OP field is 20.3 mW with resonant detuning.

Fig. 2 shows other unique characteristics. The large phase-sensitive particle number resulting from multiple amplifications even without any initial seeds, provides a measured power of ~5μW for the optical interference arm in continuous wave mode. This corresponds to $2\times10^{13}$/s photons in the optical arm and the same atomic spin in the atomic arm. A large N experimentally ensures the attainment of high absolute phase sensitivity of both optical and atomic phases, and the corresponding SQL of the phase sensitivity $\delta\varphi_{\mathrm{SQL}} = 1/\sqrt{N} = 1.6\times10^{-7}$ rad/$\sqrt{\mathrm{Hz}}$ with phase-sensitive photon number $N = 4\times10^{13}$/s. The attainment of this large number of atomic spins in the atomic arm in continuous mode is difficult since only $10^4$ atoms are present in the atom quantum interferometer[27] and ~$10^{11}$ atomic spins are in the largest atomic quantum state[33]. In our atomic vapor cell, approximately $10^{10}$ atoms are present in the $W$ beam, and this amount is not sufficient to support the generation of 4×10¹³ signal photons and atomic spins at one time. The optical pump (OP) field is essential for establishing the steady state, with the continuous pumping of the atoms in $|5^2S_{1/2}, F = 2\rangle$ back to $|5^2S_{1/2}, F = 1\rangle$. The final steady state is a result of a balance between the atomic spin generation from Raman amplification and the atomic decay from the OP field. Thus, this interferometer can be used to measure physical quantities sensitive to optical and especially atomic phases with a sensitivity of $10^{-8} \sim 10^{-7}$ rad/$\sqrt{\text{Hz}}$.

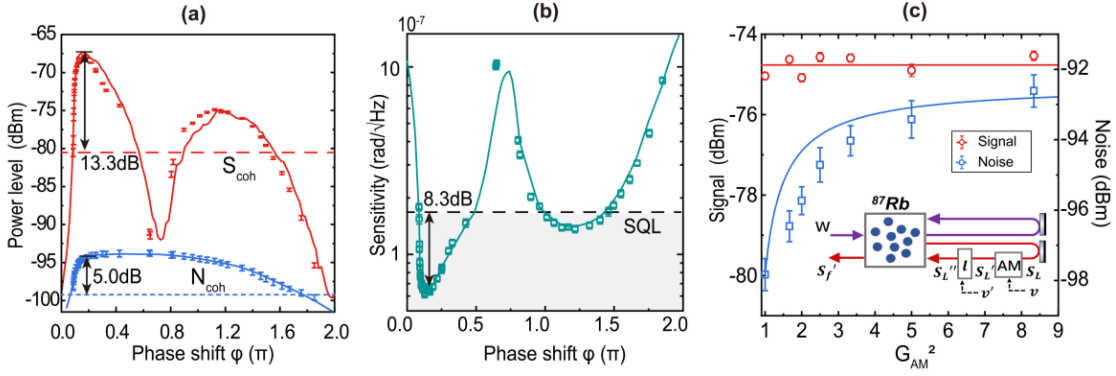

**Fig. 3 SQL-broken phase sensitivity.** In each subfigure, the symbols represent the measured data, and the solid lines are the fittings of the current interferometer. The dashed lines represent the experimental data of the laser interferometer with the same phase-sensitive particle number of N=4×10¹³/s at the optimum phase sensitivity as SQL. **(a)**, Interference signal (red solid line) and noise (blue solid line) of the current interferometer. $S_{coh}$ ($N_{coh}$): Signal (noise) power level of the laser interferometer. **(b)**, Phase sensitivity of the current sawtooth-shaped interferometer (green line and squares) along with that of the corresponding laser interferometer at the SQL (black dashed line).**(c)**, Signal (red circles) and noise (blue squares) of interference output $S_{f}'$ when the quantum correlation between the optical and atomic arms is gradually destroyed using the optical path shown in the inset. Subfigure: AM - an amplifer to amplify the field $S_L$ to $S_L'$ with amplification factor $G_{AM}^2$; $l$ - an attenuator depletes field $S_L'$ to $S_L''$ with loss rate $l$.

### SQL-broken phase sensitivity

To examine the phase sensitivity enhancement mechanism of the interferometer, its performance was measured and compared with SQL. In Fig. 3a, the sawtooth interference signal (red solid line) near the dark point on the steep side at $\varphi$=0.16$\pi$ is 13.3 dB greater than $S_{coh}$ (red dashed line). The $S_{coh}$ is the signal of cosine-shaped laser interferometer at the optimum phase sensitivity with the same number of phase-sensitive particles $N$=4×10¹³/s. One of the enhancement factors comes from quantum amplification in the backward combining process, and its gain (i.e., the ratio of the intensities of $S_f$ and $S_L$) is measured to be $I_{Sf}/I_{SL} = 2.3$, resulting in an ideal signal amplification of 7.3 dB. Therefore, the remaining slope enhancement factor of 6.0 dB, i.e., the difference between the values of 13.3 dB and 7.3 dB, is attributed to the sawtooth-wave shape caused by phase-comb. In addition, the noise level of the sawtooth interference (blue solid line) is greater than that of the coherent interference ($N_{coh}$, blue dashed line) by ~5.0 dB, which is mainly due to the internal loss in optical arm. Evidently, in our developed interferometer, the large increase in noise that is typically observed with coherent signal amplification is avoided because of the atom-light quantum correlation and phase comb superposition. Finally, as shown in Fig. 3b, the phase sensitivity of the sawtooth fringe exceeds the SQL by 8.3±0.2 dB at the largest slope point (phase shift $\varphi \sim 0.16\pi$).

Because the contribution of the atom-light quantum correlation cannot be directly evaluated under continuous wave operation, we designed an experiment using a coherent amplifier and an attenuator between forward and backward Raman amplification to analyze the respective enhancement contributions of the atom-light quantum correlation and phase comb superposition. The physical model is shown in the inset of Fig. 3c. An amplifier(AM) with an adjustable gain factor $G_{AM}$ is used to magnify the $S_L$ field to $S'_L = G_{AM}S_L + gv^\dagger$, where $v$ is the vacuum field. An attenuator $l$ is used to reduce the optical arm $S'_L$ to $S''_L = \sqrt{1-l}S'_L + \sqrt{l}v'$ with $G_{AM}\sqrt{1-l} \simeq 1$ to ensure the same intensity of $S''_L$ as the $S_L$ field at an arbitrary $G_{AM}$ by adjusting the loss $l$. Therefore, compared to the $S_L$ field, the $S''_L$ field has the same average intensity but different intensity fluctuations. Then, the $S''_L$ field enters [87]Rb atomic cell to undergo an atom-light wave-recombining process and output interference fringe $S'_f$. $S'_f$ has the same sawtooth shape as the fringe of $S_f$ in Fig. 2b. The signal and noise of the interference output $S'_f$ are shown in Fig. 3c. With increasing gain $G_{AM}^2$ from 1 to 8.5 corresponding to a loss $l$ ranging from 0 to 0.88, the signal remains unchanged at -73.8±0.2 dB because the intensities and shapes of $S'_f$ and $S_f$ are the same. However, the noise level of $S'_f$ increases from -97.8 dB to -92.6 dB because the quantum correlation between the two arms is destroyed and the excess noise of the two arms cannot be canceled when the signal is amplified. When $G_{AM}^2$=8.5, the noise increases by 5.2 dB.

Based on these experimental results, the enhancement effect of quantum correlation can be analyzed (see the Supplementary Materials.V). In the experiments, ideally, the gain of the wave-recombining process with an amplification factor of 2.3 contributes a signal enhancement of 7.3 dB. First, optical loss causes a ∼5.0 dB increase in noise and decreases the quantum enhancement to 2.3 dB. Then, amplifier AM and the attenuator $l$ further destroy the 2.3 dB quantum enhancement and even introduce excess noise from the two interference arms in the $S'_f$ output, resulting in a final increase in noise of approximately 5.2 dB at $G_{AM}^2$=8.5 in experiment. Therefore, the atom-light quantum correlation between the two arms contributes 2.3 dB, and the remaining 6.0 dB quantum enhancement results from phase comb superposition. These results directly demonstrate that the phase comb plays a decisive role in enhancing the sensitivity.

**Conclusion and discussion**

In conclusion, our developed interferometer is an atom-light hybrid quantum interferometer that benefits from dual slope enhancements via atom-light quantum correlation and phase comb superposition; its optimal phase sensitivity of both the optical and atomic phases is $\Delta\varphi = 6\times10^{-8}$ rad/$\sqrt{\text{Hz}}$, surpassing the SQL by 8.3 dB. This result is a significant step towards useful quantum interferometry; to the best of our knowledge, previously reported correlation-enhanced optical interferometry has been limited to surpassing the SQL by 4.1 dB of $N$ at the $10^{14}$ level, and correlation-enhanced atom interferometry has been limited to $10^4$ atomic spins. Further improvement of the phase sensitivity can be achieved by higher quantum breakthrough utilizing optical elements with high transmissivity and by larger $N$ using a larger beam diameter to involve more atoms. Our technology establishes a quantum correlation of up to $10^{13}$ photons and atoms in continuous mode via the Raman process and can use correlated photons to reduce the noise of atomic spins and vice versa; moreover, our technology provides a practical technique for quantum measurement of the phase-sensitive parameters, especially atomic phase sensitive parameters and can be utilized to improve the currently available ultrasensitive measurements of the optical phase-sensitive parameters, such as gravitational wave detection, gyroscopes, and biological imaging, and the parameters sensitive to the atomic phase, including gravity and magnetic fields.

## REFERENCES AND NOTES

[1]. C. W. Gardiner and P. Zoller, Quantum Noise, *Springer*, 2nd edition, (2000).

[2]. Samuel L. Braunstein, Quantum Limits on Precision Measurement of Phase, *Phys. Rev. Lett.* 69, 3598 (1992).

[3]. Vittorio Giovannetti, Seth Lloyd, and Lorenzo Maccone, Quantum metrology. *Phys. Rev. Lett.* 96, 010401 (2006).

[4]. Kim, T., Pfister, O., Holland, M. J., Noh, J., & Hall, J. L. Influence of decorrelation on Heisenberg-limited interferometry with quantum correlated photons. *Phys. Rev. A.* 57(5), 4004. (1998).

[5]. Greve, G. P., Luo, C., Wu, B., Thompson, J. K. Entanglement-enhanced matter-wave interferometry in a high-finesse cavity. *Nature*, 610(7932), 472-477 (2022).

[6]. Pan, J. W., Chen, Z. B., Lu, C. Y., Weinfurter, H., Zeilinger, A., & Żukowski, M. Multiphoton entanglement and interferometry. *Reviews of Modern Physics*, 84(2), 777 (2012).

[7]. Pezze, L., Smerzi, A., Oberthaler, M. K., Schmied, R., & Treutlein, P. Quantum metrology with nonclassical states of atomic ensembles. *Reviews of Modern Physics*, 90(3), 035005 (2018).

[8]. Ou, Z. Y., & Li, X. Quantum SU (1, 1) interferometers: Basic principles and applications. *APL Photonics*, 5(8) (2020).

[9]. B. Yurke, S.L. McCall, J.R. Klauder, SU (2) and SU (1, 1) interferometers. *Phys. Rev. A.* 33, 4033 (1986).

[10]. Sergei Slussarenko, Morgan M. Weston, Helen M. Chrzanowski, Lynden K. Shalm, Varun B. Verma, Sae Woo Nam, and Geoff J. Pryde, Unconditional violation of the shot-noise limit in photonic quantum metrology, *Nat. Photonics*, 11, 700 (2017).

[11]. Shahram Panahiyan, Carlos Sachez Munoz, Maria V. Chekhova, and Frank Schlawin, Nonlinear Interferometry for Quantum-Enhanced Measurements of Multiphoton Absorption, *Phys.Rev.Lett.* 130, 203604 (2023).

[12]. Gustavson, T. L., Landragin, A., & Kasevich, M. A. Rotation sensing with a dual atom-interferometer Sagnac gyroscope. *Classical and Quantum Gravity*, 17(12), 2385 (2000).

[13]. Durfee, D. S., Shaham, Y. K., & Kasevich, M. A. Long-term stability of an area-reversible atom-interferometer Sagnac gyroscope. *Phys.Rev.Lett.* 97(24), 240801 (2006).

[14]. Dimopoulos, S., Graham, P. W., Hogan, J. M., & Kasevich, M. A. Testing general relativity with atom interferometry. *Phys.Rev.Lett.* 98(11), 111102 (2007).

[15]. Poisson, E. Measuring black-hole parameters and testing general relativity using gravitational-wave data from space-based interferometers. *Phys.Rev.D.* 54(10), 5939 (1996).

[16]. Pitkin, M., Reid, S., Rowan, S., & Hough, J. Gravitational wave detection by interferometry (ground and space). *Living Reviews in Relativity*, 14, 1-75 (2011).

[17]. Abramovici, A., Althouse, W. E., Drever, R. W., Gürsel, Y., Kawamura, S., Raab, F. J., ... & Zucker, M. E. LIGO: The laser interferometer gravitational-wave observatory. *Science*, 256(5055), 325-333 (1992).

[18]. Abbott, B. P., Abbott, R., Abbott, T., Abernathy, M. R., Acernese, F., Ackley, K., ... & Cavalieri, R. Observation of gravitational waves from a binary black hole merger. *Phys.Rev.Lett.* 116(6), 061102 (2016).

[19]. Yepiz-Graciano, P., Ibarra-Borja, Z., Ramírez Alarcón, R., Gutiérrez-Torres, G., Cruz-Ramírez, H., Lopez-Mago, D., & U'Ren, A. B. Quantum optical coherence microscopy for bioimaging applications. *Phys.Rev.Applied*, 18(3), 034060 (2022).

[20]. Black, A. N., Nguyen, L. D., Braverman, B., Crampton, K. T., Evans, J. E., & Boyd, R. W. Quantum-enhanced phase imaging without coincidence counting. *Optica*, 10(7), 952-958. (2023).

[21]. Moreau, P. A., Toninelli, E., Gregory, T., & Padgett, M. J. Imaging with quantum states of light. *Nature Reviews Physics*, 1(6), 367-380. (2019).

[22]. Brida, G., Genovese, M., & Ruo Berchera, I. Experimental realization of sub-shot-noise quantum imaging. *Nat. Photonics*, 4(4), 227-230 (2010).

[23]. Taylor, M. A., & Bowen, W. P. Quantum metrology and its application in biology. *Physics Reports*, 615, 1-59 (2016).

[24]. Wei Du, Jia Kong, Guzhi Bao, et al., SU(2)-in-SU(1,1) Nested Interferometer for High Sensitivity, Loss-Tolerant Quantum Metrology, *Phys. Rev. Lett.* 128, 033601 (2022).

[25]. Hudelist, F., Kong, J., Liu, C., Jing, J., Ou, Z. Y., & Zhang, W. Quantum metrology with parametric amplifier-based photon correlation interferometers. *Nat.Commun*, 5(1), 3049 (2014).

[26]. Du, W., Jia, J., Chen, J. F., Ou, Z. Y., & Zhang, W. Absolute sensitivity of phase measurement in an SU (1, 1) type interferometer. *Opt. Letters*, 43(5), 1051-1054 (2018).

[27]. Mao, T. W., Liu, Q., Li, X. W., Cao, J. H., Chen, F., Xu, W. X., ... & You, L. Quantum-enhanced sensing by echoing spin-nematic squeezing in atomic bose–einstein condensate. *Nat. Physics*, 19(11), 1585-1590 (2023).

[28]. Gross, C., Zibold, T., Nicklas, E., Esteve, J., & Oberthaler, M. K. Nonlinear atom interferometer surpasses classical precision limit. *Nature*, 464(7292), 1165-1169 (2010).

[29]. Liu, Q., Wu, L. N., Cao, J. H., Mao, T. W., Li, X. W., Guo, S. F., ... & You, L. Nonlinear interferometry beyond classical limit enabled by cyclic dynamics. *Nat. Physics*, 18(2), 167-171 (2022).

[30]. Esteve, J., Gross, C., Weller, A., Giovanazzi, S., & Oberthaler, M. K. Squeezing and entanglement in a Bose–Einstein condensate. *Nature*, 455(7217), 1216-1219 (2008).

[31]. Hosten, O., Engelsen, N. J., Krishnakumar, R., & Kasevich, M. A. Measurement noise 100 times lower than the quantum-projection limit using entangled atoms. *Nature*, 529(7587), 505-508 (2016).

[32]. Colombo, S., Pedrozo-Peñafiel, E., Adiyatullin, A. F., Li, Z., Mendez, E., Shu, C., & Vuletić, V. Time-reversal-based quantum metrology with many-body entangled states. *Nat. Physics*, 18(8), 925-930 (2022).

[33]. Bao, H., Duan, J., Jin, S., Lu, X., Li, P., Qu, W., ... & Xiao, Y. Spin squeezing of 1011 atoms by prediction and retrodiction measurements. *Nature*, 581(7807), 159-163 (2020).

[34]. Cao, H., Radhakrishnan, C., Su, M., Ali, M. M., Zhang, C., Huang, Y. F., ... & Guo, G. C. Fragility of quantum correlations and coherence in a multipartite photonic system. *Phys. Rev. A*, 102(1), 012403(2020).

[35]. Anderson, Brian E., et al. "Phase sensing beyond the standard quantum limit with a variation on the SU (1, 1) interferometer." *Optica*, 4.7: 752-756 (2017).

[36]. Shimizu, A., & Miyadera, T. Stability of quantum states of finite macroscopic systems against classical noises, perturbations from environments, and local measurements. *Phys. Rev. Lett.* 89(27), 270403 (2002).

[37]. Israel, Y., Afek, I., Rosen, S., Ambar, O., & Silberberg, Y. Experimental tomography of NOON states with large photon numbers. *Phys. Rev. A*, 85(2), 022115 (2012).

[38]. Liu, Y., Li, J., Cui, L., Huo, N., Assad, S. M., Li, X., & Ou, Z. Y. Loss-tolerant quantum dense metrology with SU (1, 1) interferometer. *Opt. Express*, 26(21), 27705-27715 (2018).

[39]. Kurizki, G., Bertet, P., Kubo, Y., Mølmer, K., Petrosyan, D., Rabl, P., & Schmiedmayer, J. Quantum technologies with hybrid systems. *Proceedings of the National Academy of Sciences*, 112(13), 3866-3873 (2015).

[40]. Awschalom, D. D., Hanson, R., Wrachtrup, J., & Zhou, B. B. Quantum technologies with optically interfaced solid-state spins. *Nat. Photonics*, 12(9), 516-527(2018).

[41]. Chen, B., Qiu, C., Chen, S., Guo, J., Chen, L. Q., Ou, Z. Y., & Zhang, W. Atom-light hybrid interferometer. *Phys. Rev. Lett.* 115(4), 043602 (2015).

[42]. Manceau, M., Leuchs, G., Khalili, F., & Chekhova, M. Detection loss tolerant supersensitive phase measurement with an SU (1, 1) interferometer. *Phys. Rev. Lett.* 119(22), 223604 (2017).

[43]. Frascella, G., Agne, S., Khalili, F. Y., & Chekhova, M. V. Overcoming detection loss and noise in squeezing-based optical sensing. *npj Quantum Inf*, 7(1), 72 (2021).

[44]. J. Gough and M. R. James, The Series Product and Its Application to Quantum Feedforward and Feedback Networks, *IEEE Trans.* Autom. Control 54(11), 2530–2544 (2009).

[45]. J. Gough and M. R. James, Quantum Feedback Networks: Hamiltonian Formulation, *Commun. Math. Phys.* 287(3), 1109–1132 (2009).

[46]. J. Combes, J. Kerckhoff, and M. Sarovar, The SLH framework for modeling quantum input-output networks, *Adv. Phys.*: X 2(3), 784–788 (2017).

[47]. Zhifei Yu, Bo Fang, Pan Liu, Shuying Chen, Guzhi Bao, Chun-Hua Yuan, and L. Q. Chen. Sensing the performance enhancement via asymmetric gain optimization in the atom-light hybrid interferometer. *Opt. Express*, 30(7), 11514-11523 (2022).

## ACKNOWLEDGMENTS

**Funding**
National Natural Science Foundation of China Grants No. U23A207, No. 12274132, No.11974111, No. 12364046, No. 11964014.
Innovation Program of Shanghai Municipal Education Commission No. 202101070008E00099.
Major Discipline Academic and Technical Leader Training Program of Jiangxi Province Grant No. 20204BCJ23026.
China Postdoctoral Science Foundation Grants No. 2023M741187, No. GZC20230815.
Fundamental Research Funds for the Central Universities.

**Data availability**
The datasets generated and analysed during this study are available from the corresponding authors upon reasonable request.

**Code availability**
The codes that support the findings of this study are available from the corresponding author upon reasonable request.

**Author contributions**
L.C. supervised the project and conceived the idea. W.H., X.L., and L.C. designed the experiment, performed the measurements, and analysed the data together with all other authors. J.Z., Z.W. and Y.W. provided support in experimental techniques. B.F., K.Z., C.Y.,and W.Z. provided theoretical support. W.H., X.L., B.F., and L.C. wrote the manuscript. All authors discussed the experiment implementation and results and contributed to the manuscript.

**Competing interests**
The authors declare no competing interests.

## SUPPLEMENTARY MATERIALS

Supplementary Text

Figs. S1 to S3

References (44–47)